\documentclass[twocolumn,showpacs,superscriptaddress]{revtex4}
\usepackage{graphicx}
\usepackage{dcolumn}
\usepackage{bm}
\usepackage{amsmath}

\newcommand{\be}{\begin{equation}}
\newcommand{\ee}{\end{equation}}

\begin{document}

\title{Two-dimensional solitons with hidden and explicit vorticity in bimodal cubic-quintic media}
\author{A.S.~Desyatnikov}
\affiliation{Nonlinear Physics Centre and Centre for Ultra-high bandwidth Devices for Optical Systems (CUDOS), Research School of Physical Sciences and Engineering, Australian National University, Canberra ACT 0200, Australia} \affiliation{Institute of Applied
Physics, Westf\"{a}lische Wilhelms-Universit\"{a}t M\"{u}nster, D-48149 M\"{u}nster, Germany}
\author{D.~Mihalache}
\author{D.~Mazilu}
\affiliation{Department of Theoretical Physics, Institute of Atomic Physics, POBox MG-6,
Bucharest, Romania}
\affiliation{Institute of Solid State Theory and Theoretical Optics, Friedrich-Schiller
Universit\"{a}t Jena, Max-Wien-Platz 1, D-07743 Jena, Germany}
\author{B.A.~Malomed}
\affiliation{Department of Interdisciplinary Studies, School of Electrical Engineering,
Faculty of Engineering, Tel Aviv University, Tel Aviv 69978,Israel}
\author{C.~Denz}
\affiliation{Institute of Applied Physics, Westf\"{a}lische Wilhelms-Universit\"{a}t M\"{u}nster, D-48149 M\"{u}nster, Germany}
\author{F.~Lederer}
\affiliation{Institute of Solid State Theory and Theoretical Optics, Friedrich-Schiller
Universit\"{a}t Jena, Max-Wien-Platz 1, D-07743 Jena, Germany}

\begin{abstract}
We demonstrate that two-dimensional two-component bright solitons of an annular shape, carrying vorticities $(m,\pm m)$ in the components, may be stable in media with the cubic-quintic nonlinearity, including the \textit{hidden-vorticity} (HV) solitons of the type $(m,-m)$, whose net vorticity is zero. Stability regions for the vortices of both $(m,\pm m)$ types are identified for $m=1$, $2$, and $3$, by dint of the calculation of
stability eigenvalues, and in direct simulations. A novel feature found in the study of the HV solitons is that their stability intervals never reach the (cutoff) point at which the bright vortex carries over into a dark one, hence dark HV solitons can never be stable, contrarily to the bright ones. In addition to the well-known symmetry-breaking (\textit{external}) instability, which splits the ring soliton into a set of fragments flying away in tangential directions, we report two new scenarios of the development of weak instabilities specific to the HV solitons. One features \textit{charge flipping}, with the two components exchanging the angular momentum and periodically reversing the
sign of their spins. The composite soliton does not split in this case, therefore we identify such instability as an \textit{intrinsic} one. Eventually, the soliton splits, as weak radiation loss drives it across the border of the ordinary strong (external) instability. Another scenario proceeds through separation of the vortex cores in the two components, each individual core moving toward the outer edge of the annular soliton. After expulsion of the cores, there remains a zero-vorticity breather with persistent internal vibrations.
\end{abstract}

\maketitle

\section{Introduction}

Optical vortices are two- or three-dimensional (2D or 3D) dark or bright solitons with an embedded phase dislocation, which lends them a topological charge (``spin''). As topologically nontrivial self-trapped states, the vortices have attracted a great deal of attention, which was additionally enhanced by a potential which the 2D ones may have as reconfigurable conduits for weak optical signals \cite{book,opn02,Pramana}. Vortex solitons of the bright type are of special interest in both respects. First, being relatively compact objects, they make it possible to realize sophisticated multi-vortex configurations. On the other hand, experimental creation of bright solitons is hampered by the fact that, in media with the simplest collapse-free nonlinearities (quadratic or
saturable), they are subject to strong azimuthal instability, which splits them into a set of ordinary (zero-spin) solitons \cite{dima,Petrov}. Nevertheless, as it was first observed in direct simulations of a model with the cubic-quintic (CQ) nonlinearity in Ref.~\cite{Quiroga}, and later investigated in detail by more accurate methods (see a review in Ref.~\cite{Pramana}), bright vortices with the spin $m=1$ \cite{Quiroga}, $m=2$ \cite{Pramana}, and $m\geq 3$ \cite{Bob} may be stable if the model features competing self-focusing and self-defocusing nonlinearities. Another example which corroborates this conclusion is a model combining quadratic and self-defocusing cubic nonlinearities \cite{Sammut}. Later, stable two-component (\textit{vectorial}) solitons, carrying the vorticity set $(m,m)$, have been identified in a vectorial version of the CQ model
\cite{vectorial}.

It is relevant to mention that patterns of a similar type, in the form of ``soliton necklaces'', i.e., ring-shaped chains of the fundamental ($m=0$) solitons, were introduced and studied in the model with the Kerr (cubic) nonlinearity \cite{moti}. The necklaces may be constructed with zero or non-integer angular momentum; however, they are not stationary objects, as they spread out and eventually disintegrate. A possibility to (practically) stabilize necklace-like patterns is to introduce a vectorial interaction with a fundamental soliton, which helps to support vortex-mode, dipole-mode, and multipole vector solitons \cite{josab}. Additional stabilization of two-component necklaces in vectorial model allows to construct stationary necklace-ring vector solitons \cite{necklring}, having azimuthally modulated (necklace-type) components whose densities sum up into an azimuthally uniform distribution of the total intensity. However, multipole and necklace-ring vector solitons are subject to an azimuthal instability, except of the dipole-mode vector soliton \cite{dipole} and vortex-mode soliton close to a bifurcation \cite{pelinovsky}. A particular class of such (generally,
unstable) solutions is a vector vortex soliton with equal amplitude distributions in both components \cite{Bigelow}.

A challenging issue is a possibility of the existence of stable vectorial solitons of the $(m,-m)$ type, that would feature the same annular shape (with a hole in the center) as the bright scalar or vectorial vortices with the spins, respectively, $m$ or $(m,m)$, but with zero net spin. The possibility of the existence of such objects is obvious if the coupling between the components is of the XPM (cross-phase-modulation) type, i.e., insensitive to their relative phase -- then, there is no difference in the shape
between stationary vortex solitons of the $(m,m)$ and $(m,-m)$ types. However, in the simplest collapse-free model with saturable nonlinearity, which is based on XPM-coupled nonlinear Schr\"{o}dinger (NLS) equations for local slowly varying amplitudes $E_{1,2}$, with terms like $E_{1,2}/\left(1+|E_1|^2+|E_2|^2\right) $, the $(m,-m)$ vortices are unstable [as well as their $(m,m)$ counterparts], although it was demonstrated that the instability may be partly suppressed \cite{Bigelow}. A stabilizing effect of the incoherent interaction of counter-rotating vortices was also demonstrated in an anisotropic photorefractive self-defocusing medium \cite{Mamaev}.

Compound vortices with the spin components $(1,-1,0)$ and $(1,1,2)$ were also studied in the three-wave model of the type-II second-harmonic generation, with two components of the fundamental wave, and one second-harmonic component \cite{Torres,Leblond}. It was shown that, in this model per se, vortices of both types are unstable -- against fusion into an ordinary zero-spin soliton, or splitting, respectively. The addition of a stabilizing repulsive cubic interaction makes the life expectancy of the vortices much longer, but no case of complete stabilization has been found \cite{Leblond}.

In this work, we demonstrate that, in contrast with all the previously studied models, in the two-component CQ model the $(m,-m)$ vortex solitons are \emph{rigorously stable} in a certain parameter region. In fact, this result opens up a new class of stable 2D solitons with {\it hidden vorticity} (HV). 

We start with a general two-component CQ model in the rescaled form, which describes spatial evolution (along the propagation coordinate $z$) of the light beams in a bulk medium \cite{Maimistov}, 
\begin{eqnarray}
&&i\partial_z E_{1,2}+\Delta E_{1,2}+\left( |E_{1,2}|^2+\alpha |E_{2,1}|^2\right) E_{1,2}  \notag  \label{model} \\
&&-\left( |E_{1,2}|^4+2\beta |E_1|^2|E_2|^2+\beta|E_{2,1}|^4\right) E_{1,2}=0,  \label{E12}
\end{eqnarray}
where $E_{1,2}(x,y,z)$ are the local slowly varying amplitudes of the two waves, and the Laplacian $\Delta$ is the diffraction operator acting on the transverse coordinates $(x,y)$. The real parameters $\alpha$ and $\beta$ are the cubic and quintic XPM coefficients, respectively. 

The CQ nonlinearity was experimentally observed in isotropic media, such as glasses \cite{Dijon} and organic materials \cite{China}; these materials also feature nonlinear loss, but the consideration of the conservative model is justified, as the characteristic soliton's length can be made essentially smaller than the absorption length, or the loss may be compensated by gain. In that case, $E_1$ and $E_2$ may be realized as orthogonally polarized waves, with $\alpha =2/3$ for linear and $\alpha =2$ for circular polarizations, respectively. The latter case pertains as well to a set of two waves with different carrier wavelengths. Besides that, the model (\ref{E12}) may serve as a crude isotropic approximation for the description of photorefractive media in the low-saturation regime, which is characterized by equal strengths of the XPM and SPM (self-phase-modulation) nonlinearities, $\alpha =\beta =1$ \cite{opn02}. Thus, different values of $\alpha $ and $\beta $ are physically relevant.

It is well known that fundamental ($m=0$) solitons, both scalar and vectorial [the latter ones, of the type $(0,0)$], are stable in models with any saturable nonlinearity, including the CQ model (for the isotropic scalar saturable model, this was demonstrated
more than a decade ago \cite{Enns}). Recently, stability of a vectorial soliton which is fundamental ($m=0$) in one component, and carries the vorticity $m=1$ in the other, i.e., a ``vortex-mode soliton" of the $(0,1)$ type, has been predicted close to a bifurcation in the saturable medium \cite{pelinovsky}. More complex vectorial solitons, combining a zero-vorticity configuration in one component and a topologically charged one in the other, are also possible in saturable media \cite{josab}. Such solutions definitely exist in the CQ model too, and they are plausibly stable in a broad parametric area. 

\section{Stationary solutions and stability analysis}

\begin{figure}
\includegraphics[width=80mm]{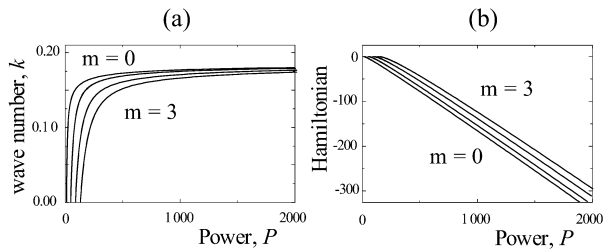}
\caption{The wavenumber $k$ (a) and Hamiltonian (b), calculated as per Eq.~(\ref{H-radial}), versus total power, $P=P_1+P_2$, for the vectorial vortex solitons of the $(m,\pm m)$ types.}
\label{fig1}
\end{figure}

In this work we focus on the vectorial vortex solitons of the $(m,\pm m)$ types with symmetric components, which carry equal wave numbers $k $ (non-symmetric solutions with $k_1\neq k_2$ are possible too, but we do not expect their properties to be drastically different from those of the symmetric solitons):
\begin{equation} 
\left(
\begin{array}{l}E_1\\E_2\end{array}\right) =V(r)\exp (ikz)\left(
\begin{array}{l}\exp (im\varphi ) \\ \exp (\pm im\varphi )\end{array}\right),  
\label{V}
\end{equation}
where $r$ and $\varphi $ are the polar coordinates in the plane $(x,y)$, and the real function $V$ obeys the equation 
\begin{equation}
kV=\hat{D}_mV+\left( 1+\alpha \right) V^3-(1+3\beta )V^5,  
\label{ODE}
\end{equation}
where $\hat{D}_m\equiv d^2/dr^2+r^{-1}d/dr-m^2r^{-2}$. Using the transformation $V=R(1+\alpha)^{1/2}(1+3\beta)^{-1/2}$, $r\rightarrow r(1+\alpha )^{-1}(1+3\beta )^{1/2}$, and $k\rightarrow k(1+\alpha)^{-2}(1+3\beta)$, we cast Eq. (\ref{ODE}) in the form 
\begin{equation}
kR=\hat{D}_mR+R^3-R^5,  \label{simpleR}
\end{equation}
which is supplemented by the boundary condition $R\sim r^{|m|}$ at $r\rightarrow 0$. For $r\rightarrow \infty $, there are two types of solutions to Eq.~(\ref{simpleR}), \emph{coexisting} in the medium with competing nonlinearities \cite{dark}, viz., bright solitons with $R\sim \exp \left(-\sqrt{k}r\right) /\sqrt{r}$, and dark solitons with $R^2(r=\infty)=(1+\sqrt{1-4k})/2$. Integral characteristics of the bright vectorial soliton are its partial powers in both components, 
\begin{equation}
P_{1,2}\equiv \frac{2\pi}{1+\alpha}\int_0^\infty rdrR^2(r).  
\label{P}
\end{equation}
Global characteristic of the soliton families, in the form of dependencies $k(P)$, are displayed in Fig.~\ref{fig1}(a) for $\alpha=\beta=1$, here the total power $P=P_1+P_2$. The \textit{cutoff} (largest possible) value of $k$ for the bright-soliton family is the same as for the family of commonly known 1D solitons in the CQ model, which is $k\left( P=\infty \right) =3/16\equiv 0.1875$; at this value of $k$, the bright solitons become infinitely broad (approaching a finite maximum amplitude, $R_{\max}=\sqrt{3}/2$), i.e., they go over into dark solitons. We note that the Vakhitov-Kolokolov criterion, $dk/dP>0$, which is a necessary condition for stability of the solitons \cite{VK}, is satisfied for all these solutions. Actually, it guarantees the stability against radial perturbations, but not against symmetry-breaking ones, which are known to be most dangerous for the stability of vortex solitons \cite{Pramana}.

Note that the partial angular momenta of the components are not conserved independently, the corresponding integral of motion being the total angular momentum, $M=M_1(z)+M_2(z)$, where the initial partial momenta 
\begin{equation}
M_{1,2}=\int \mathrm{Im}\left( E_{1,2}^\ast\frac{\partial E_{1,2}}{\partial \varphi }\right) d\mathbf{r}  
\label{M}
\end{equation}
(recall $\varphi $ is the angular coordinate). Two distinct values of total angular momentum, $M=2mP$ and $M=0$, correspond to two types of solutions, $(m,m)$ and $(m,-m)$. Obviously, both solitons coincide in shape, therefore the diagrams for them shown
in Fig.~\ref{fig1} are identical. However, \emph{stability} of the two types of the solutions may be completely different. An attempt to describe such a difference was made, based on the ``thin-ring" approximation, in a saturable medium \cite{Bigelow}; however, all the vortices are unstable in that medium. The ``thin-ring" approximation definitely does not apply to the ``wide-ring" vortex solitons studied here.

The most important results are adequately represented by the case of $\alpha =\beta =1$, on which we focus below. In this case, the Hamiltonian of Eqs.~(\ref{E12}) is \begin{equation}
H=\int \left( |\nabla E_1|^2+|\nabla E_2|^2-\tfrac{1}{2}I^2+\tfrac{1}{3}I^3\right) d\mathbf{r}  
\label{H}
\end{equation}
with the total intensity $I=|E_1|^2+|E_2|^2$. The transformation of variables which leads to the normalized equation~(\ref{simpleR}) amounts to $V(r)\equiv R(r)/\sqrt{2}$, so that $I=R^2$. With the latter substitution, Eq.~(\ref{H}) reduces to the Hamiltonian of a scalar vortex with charge $m$,
\begin{equation}
H=2\pi \int_0^\infty\left[ \left( \frac{dR}{dr}\right)^2+\frac{m^2}{r^2}R^2-\frac{1}{2}R^2+\frac{1}{3}R^3\right]rdr.  \label{H-radial}
\end{equation}

Perturbed vortex-soliton solutions are sought for in the form [cf. Eq.~(\ref{V}) for the unperturbed ones] 
\begin{eqnarray}
E_1=\exp (ikz+im\varphi) [R(r)/\sqrt{2}+\tilde{f}+\tilde{g}^\ast],\notag \\
E_2=\exp (ikz\pm im\varphi) [R(r)/\sqrt{2}+\tilde{p}+\tilde{q}^\ast],
\label{perturbed}
\end{eqnarray}
here $\{\tilde{f},\tilde{g},\tilde{p},\tilde{q}\}\equiv\{f(r),g(r),p(r),q(r)\}\exp(\lambda_s z+is\varphi)$ with complex eigenvalue $\lambda_s$ and an arbitrary integer azimuthal index $s$. Substitution of these expressions into linearized equations (\ref{model}) yields a system 
\begin{equation}
i\lambda_s\left(\begin{array}{c}f\\g\\p\\q\end{array}\right)=\left[
\begin{array}{llll}
\;\;\;\hat{L}^+&\;\;\;A&\;\;\;A&\;\;\;A\\
-A & -\hat{L}^-&-A&-A\\
\;\;\;A&\;\;\;A&\;\;\;\hat{L}^\pm&\;\;\;A\\
-A & -A & -A & -\hat{L}^\mp\end{array}\right]
\left(\begin{array}{c}f\\g\\p\\q\end{array}\right),  
\label{linear}
\end{equation}
where $\hat{L}^\pm\equiv \hat{D}_{m\pm s}-k+R^2\left(3/2-2R^2\right)$ and $A\equiv R^2\left(1/2-R^2\right) $. The signs $\pm$ in Eq. (\ref{linear}) correspond to the two states $(m,\pm m)$. 

Note that in the case of $(m,m)$ solutions, the matrix in Eqs.~(\ref{linear}) has a block $[2\times 2]$ structure, hence the eigenmodes degenerate ($f=p$ and $g=q$), and the linear stability problem reduces to the one for the scalar vortex, cf. Refs.~\cite{Pramana, Bob}. Therefore, the stability properties of the $(m,m)$ vectorial vortices are \textit{completely identical} to those of their scalar counterparts. However, the degeneracy does not take place for the HV solitons of the $(m,-m)$ type,
which clearly shows the difference in the stability problem for the two types of vectorial vortex solitons. 

\begin{figure}
\includegraphics[width=80mm]{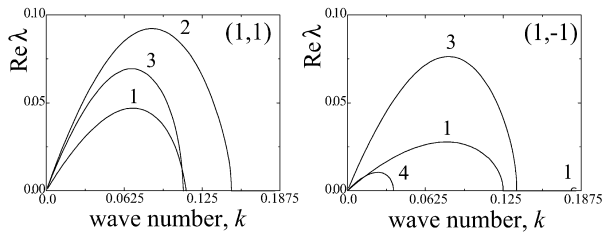}
\caption{Growth rates of perturbation eigenmodes with different values of the azimuthal index $s$ (indicated next to the curves) for the vectorial vortex solitons of the types $(1,\pm 1)$.}
\label{fig2}
\end{figure}

Stability eigenvalues were found from numerical solution of Eqs.~(\ref{linear}). In Fig.~\ref{fig2}, we display dependencies of the eigenvalues with different values of the azimuthal index $s$ on the wavenumber $k$ for the vectorial vortex solitons of the $(1,1)$ and $(1,-1)$ types. The maximum growth rate is found for the modes with, respectively, $s=2$ and $s=3$. 

\begin{figure}
\includegraphics[width=80mm]{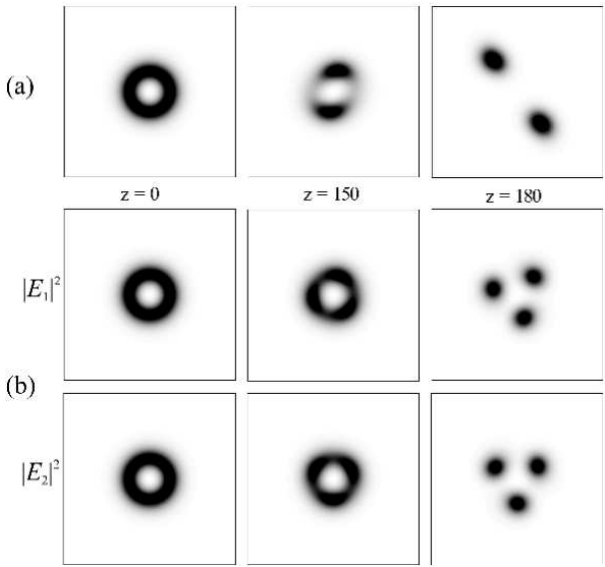}
\caption{Instability-induced evolution of the vectorial vortex solitons with $k=0.1$. (a) One of two identical components of the $(+1,+1)$ soliton. (b) Two components of the $(+1,-1)$ soliton with implicit vorticity.}
\label{fig3}
\end{figure}

The above results comply with direct simulations of the evolution of the vortices shown in Fig.~\ref{fig3} for $k=0.1$, when the linear stability analysis predicts that the solitons of both types are unstable. The symmetry-breaking instability modes of the $(1,1)$ soliton in both components are identical, therefore in the Fig.~\ref{fig3}(a) we display only one of them. The observed dynamics of two splinters (which are zero-vorticity vector solitons), generated from this soliton, is exactly the same as was observed for the scalar vortex: the splinters fly away in tangential directions \cite{dima,Bob}.

A totally different scenario is observed in Fig.~\ref{fig3}~(b), where, in each component, three splinters of the initial HV soliton start to move in radial directions (cf. similar observations in Refs.~\cite{necklring,dips}). At this stage of the HV-soliton's break-up, as it seen in the panel corresponding to $z=180$ in Fig. \ref{fig3}~(b), the triangular sets of the splinters in the two components are slightly misaligned. With the further propagation, the separation of the splinters ceases, and they eventually \emph{fuse} into a spinless $(0,0)$ vectorial soliton.

\begin{figure}
\includegraphics[width=80mm]{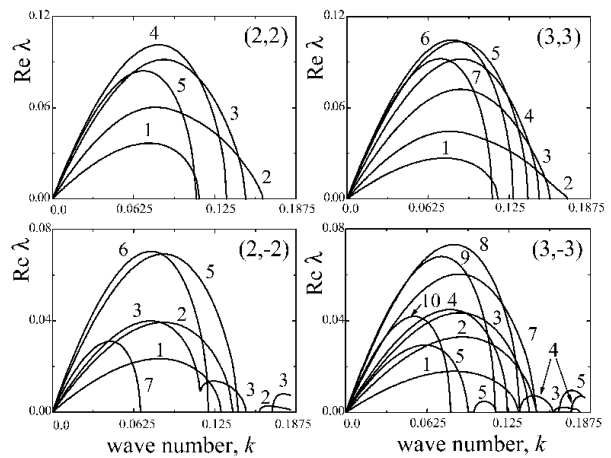}
\caption{The same as in Fig.~\ref{fig2} for the vectorial vortex solitons of the types $(m,\pm m)$ for $m=2,3$.}
\label{fig4}
\end{figure}

Figure~\ref{fig4} displays the instability growth rates found from Eq. (\ref{linear}) for higher-order vortex solitons, of the $(m,\pm m)$ types, for $m=2$ and $3$. As is seen from these figures, each type of the soliton has its stability area, as summarized in Table I. From these results, we conclude that, for all the solitons with explicit vorticity [the $(m,m)$ type], the stability regions extend up to the cutoff value, $k=0.1875$, which implies that these solitons continuously carry over into stable vortices of the dark-soliton type, similar to what is known is the scalar case. On the contrary, for the HV solitons [the $(m,-m)$ type], the stability interval \emph{never reaches} the cutoff value, i.e., dark vortices of the same type are \emph{always unstable}. In fact, stability intervals for vortex solitons terminating at $k$ smaller than the cutoff value (``stability islands") have never been reported before. In the case of the higher-order solitons (with $m=2$ and $3$), the $(m,-m)$ HV solitons have a smaller stability domain than their $(m,m)$ counterparts. It is noteworthy too that the azimuthal index $s$ of the most unstable eigenmode depends on the type of the soliton,
and for the higher-order ones, of the types $(2,\pm 2)$ and $(3,\pm 3)$, the most dangerous value of $s$ depends on $k$ too. 

\section{Charge flipping}

\begin{figure}
\includegraphics[width=80mm]{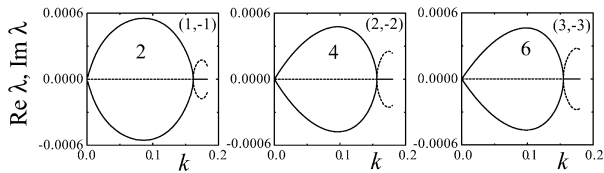}
\caption{The growth rate of the small ``internal" instability eigenmodes for the hidden-vorticity solitons of the type $(m,-m)$ with $m=1$,$2$, and $3$, the corresponding azimuthal indices being $s=2m$.}
\label{fig5}
\end{figure}

Performing numerical analysis of the linear-stability problem, based on Eq.~(\ref{linear}), with higher accuracy we have found additional very small unstable eigenvalues for even azimuthal indices ($s=2m$), with the growth rate $\simeq 5\times 10^{-4}$, which were not visible in Figs.~\ref{fig2} and \ref{fig4} (but which definitely exceed a possible numerical error). In Fig.~\ref{fig5}, we plot bifurcations which give rise to these miniscule eigenvalues (which are ordinary bifurcations of the pitchfork type), for the solitons with $m=1$,$2$, and $3$. As a result of these additional very weak perturbations (which may be unobservable in the experiment), the corresponding rigorously defined stability regions for the $(m,-m)$ HV solitons are considerably reduced with respect to their $(m,m)$ counterparts, see Table I for a summary of the output of the stability calculations.

\begin{figure}
\includegraphics[width=80mm]{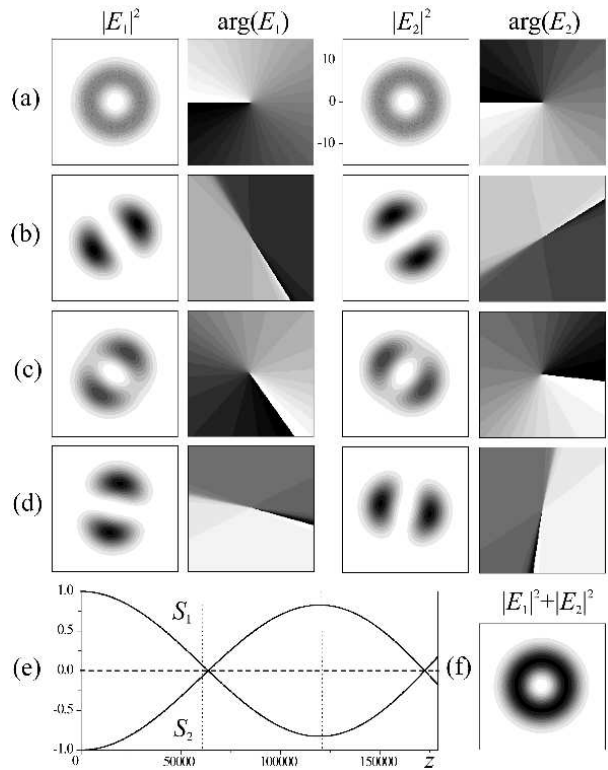}
\caption{Evolution of the ``internally unstable" vortex soliton with the hidden vorticity, of the $(1,-1)$ type, for $k=0.14$. The propagation distance is $z=0$ (a), $60000$ (b), $120000$ (c), and 175000 (d). The dynamical exchange of the spins $S_{1,2}$ (solid lines in (e)) is accompanied by \textit{charge flipping} (sign reversal of the vorticity in each component), as clearly visible in the phase diagrams for $\mathrm{arg}(E_{1,2})$ in (a)-(d). The total spin $S=(S_1+S_2)/2$ keeps its zero value (the dashed line in (e)). The net intensity, $|E_1|^2+|E_2|^2$, keeps its azimuthal homogeneity in the course of the entire propagation, as shown in (f) for $z=175000$. Vertical dotted lines correspond to snapshots (b) and (c).}
\label{fig6}
\end{figure}

To visualize the development of the weak instability revealed by Fig.~\ref{fig5}, we simulated the propagation of the solitons of both types $(1,\pm 1)$ with $k=0.14$, adding $10$\% of the random noise. For the solitons with the explicit vorticity, the instability growth rate is $\mathrm{Re}\lambda =0.0373$, and the dynamics follows the ``usual" break-up scenario shown in Fig.~\ref{fig4} (a), therefore we do not display it again. The growth rate of the same instability mode (with $s=2$) but for the HV soliton of the $(1,-1)$ type is two orders of magnitude smaller,
$\mathrm{Re}\lambda \approx 0.00036$, therefore, noticeable development of the instability should be expected after having passed the distance $\sim 10^{4}$. Although so large propagation distances can hardly be achieved experimentally (in the experiment, these solitons will seem as completely stable ones), the issue is of principal interest, therefore we have completed the numerical analysis and found the soliton's dynamics of a novel kind. The results are summarized in Fig.~\ref{fig6}, where we show the intensity and phase distributions for both components up to $z=175000$.

To explain the complex internal dynamics observed in Fig.~\ref{fig6}, we recall the theory developed for necklace-ring vector solitons in Ref.~\cite{necklring}. The HV solutions that we consider here belong to a particular branch of a much broader class of
necklace-ring vectorial solitons, namely, the ones with equal powers in both components, $P_1=P_2=P/2$. In the most general case, these solutions can be represented as [cf. Eq.~(\ref{V})] \begin{equation}
\left(\begin{array}{l}E_1\\E_2\end{array}\right)=R(r)e^{ikz}\mathbf{\Theta \Psi \Phi
}\left(\begin{array}{l}\cos \left( m\varphi \right)  \\
\sin \left( m\varphi \right) \end{array}\right).
\label{Ematr}
\end{equation}
Here the matrix 
\begin{equation}
\mathbf{\Theta}=\left[\begin{array}{lr}
e^{i\theta_1} & 0 \\0 & e^{i\theta_2}\end{array}\right],  
\label{Tmatr}
\end{equation}
with arbitrary constants $\theta_{1,2}$ indicates the phase invariance of the solutions, which is the symmetry property amenable for the conservation of the partial powers $P_{1,2}$. The matrix $\mathbf{\Phi}$ is simply a rotational transformation in the transverse plane, 
\begin{equation}
\mathbf{\Phi}=\left[\begin{array}{rr}
\cos \left( m\varphi_0\right)  & \ \sin \left( m\varphi_0\right)  \\
-\sin \left( m\varphi_0\right)  & \ \cos \left( m\varphi_0\right)
\end{array}\right],  
\label{phi}
\end{equation}
where the arbitrary constant angle $\varphi_0$ reflects the rotational invariance responsible for the conservation of the total angular momentum. Note that, for radially symmetric solutions Eq.~(\ref{V}), the transformation $\varphi \rightarrow \varphi -\varphi_0$, described by Eq.~(\ref{phi}), is equivalent to the phase shift accounted for by Eq.~(\ref{Tmatr}) with $\theta_{1,2}=\mp m\varphi_0$.

With the accuracy of arbitrary phase shift being already absorbed by the matrix $\mathbf{\Theta}$, the linear transformation $\mathbf{\Psi}$ in Eq.~(\ref{Ematr}) is given by
\begin{equation}
\mathbf{\Psi}=\left[ \begin{array}{rr}
\cos \psi  & \ i\sin \psi  \\ \sin \psi  & \ \pm i\cos \psi \end{array}\right],  \label{psi}
\end{equation}
where the constant parameter $\psi$ describes the rotation in the space of the components $(E_1,E_2)$, similar to the Manakov system, and corresponds to the conservation if the ``isotopic spin'', $i\int \{E_1E_2^\ast-E_1^\ast E_2\}d\mathbf{r}$. For any value of $\psi $, the latter expression is zero in our case. In addition, the parameter $\psi$ uniquely defines the initial values of the partial spins $S_i=M_i/P_i$ [recall the partial angular momenta were defined in Eq.~(\ref{M})]:
\begin{equation} 
S_1=m\sin (2\psi),\;S_2=\pm S_1,
\label{S}
\end{equation}
and, therefore, it determines the total spin (dynamical invariant), $S\equiv M/P=\tfrac{1}{2}(S_1+S_2)=\tfrac{1}{2}(m\pm m)\sin (2\psi)$.

Among possible stationary solutions conforming to Eq.~(\ref{Ematr}) are those with zero, fractional ($0<S<m$), and integer total spin ($S=m$). We focus here on two cases which correspond to Eq.~(\ref{V}): the $(m,m)$ type of the solutions, with the total angular momentum attaining its maximum possible value, $M=mP$ (i.e., $S=m$), and the HV solutions of $(m,-m)$ type, with $M=S=0$. Both of them represent radially symmetric vector vortices with $|\psi|=\pi/4$; the transformations $\psi\rightarrow -\psi$, $m\rightarrow -m$, and $(m,\pm m)\rightarrow (-m,\mp m)$ are all equivalent.

Solutions with the explicit vorticity, $S=m$, correspond to the upper sign in Eq.~(\ref{psi}). The partial spins assume the maximum possible values in this case, $S_{1,2}=m$, thus the only corresponding configuration is the one with $|\psi |\equiv \pi /4$, and the vortex soliton of this type always has axially symmetric (ring-shaped) components. In addition, the exchange of angular momentum between components is forbidden in this case. 

The HV solutions with $S=0$ are drastically different. They correspond to the lower sign in Eq.~(\ref{psi}) for an \textit{arbitrary} value of $\psi $. These include the HV vectorial solitons for $\psi =\pm \pi/4$, and also solutions with intensity distributions in the two components in the form of two \textit{crossed multipoles}, for $\psi =0$ and $S_{1,2}=0$. Similar solution with $m=1$, or a dipole-dipole vectorial soliton, was investigated theoretically and experimentally in Refs.~\cite{necklring, dips} and found to be azimuthally unstable in saturable medium. The $\psi$-values from the interval $-\pi /4<\psi <\pi /4$ determine the \textit{depth of the azimuthal modulation} in each of the two components, that sum up into the azimuthally uniform distribution of the total intensity $I=R^2$ \cite{necklring}. Thus, the continuous soliton family includes crossed multipoles with different values of the azimuthal modulation depth and opposite fractional values of the partial spins. It is important to note that, because all the dynamical invariants do not depend on $\psi$, for the whole branch of the
HV vectorial solitons, parametrized by $\psi$, the dynamical exchange of the angular momentum between components is possible.

\begin{figure}
\includegraphics[width=70mm]{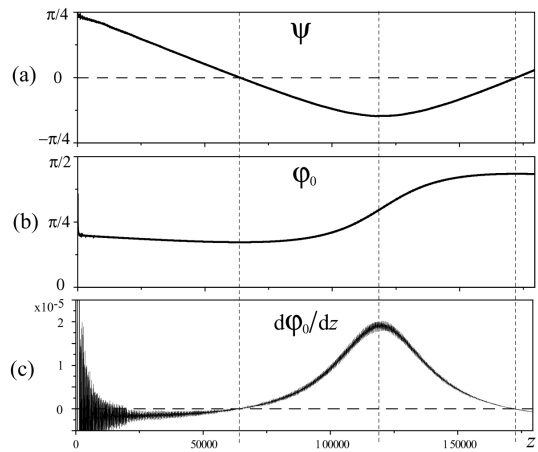}
\caption{Evolution of the parameters $\psi$ from Eq.~(\ref{psi}) in (a) and $\varphi_0$ from Eq.~(\ref{phi}) in (b), corresponding to the adiabatic ``sliding" across the soliton family, defined by Eq.~(\ref{Ematr}) in the course of the evolution displayed in Fig.~\ref{fig6}. In (c), the angular velocity of the internal rotation is shown; strong numerical noise on this curve reflects the uncertainty of the value of the parameter
$\varphi_0$ (which is arbitrary, in the general case). Vertical dashed lines indicate the charge-flipping (spin-reversal) points, close to those shown in Figs.~\ref{fig6} (b) and (d).}
\label{fig7}
\end{figure}

Using the decomposition in Eq.~(\ref{Ematr}), we can now explain the dynamics in Fig.~\ref{fig6} in terms of a slow evolution of the parameters $\varphi_0$ and $\psi $, as shown in Fig.~\ref{fig7}. In other words, instead of the modulational instability along the azimuthal direction, that would lead to fragmentation of the initial annular soliton, the instability modes from Fig.~\ref{fig5} initiate slow sliding of the solution
across the continuous manifold with the independent parameters $\varphi_0$ and $\psi $. At each stage of the evolution, such as those corresponding to the frames (a)-(d) in Fig.~\ref{fig6}, we observe a slightly perturbed stationary solution with varying
$\varphi_0$ and $\psi$ (the arbitrary phases $\theta_{1,2}$ are of no importance because the XPM interaction between the components is phase-insensitive).

Indeed, the random noise on the level of $10$\% of the soliton's amplitude, added to the HV soliton in Fig.~\ref{fig6}(a), quickly dissipates, and, in the course of the first several thousands units, the vector HV soliton propagates without any noticeable
change. Then, as is seen in Fig.~\ref{fig7}(a), the parameter $\psi$ decreases and the components assume a shape of two crossed dipoles. When the modulation depth reaches its maximum for $\psi=0$, the solution is, simply, $\{E_1,E_2\}=R(r)\{\cos (\varphi -\varphi_0),\sin (\varphi -\varphi_0)\}$, hence, at this point, \emph{neither component} contains any vorticity, as is indeed seen in Fig.~\ref{fig6}(b). Because the parameter $\varphi_0$ is arbitrary, its particular value at this stage (Fig.~\ref{fig7} (b)) depends on the initial noise.

With the further propagation, the components $E_{1,2}$ almost restore their initial annular shape and develop the phase dislocations corresponding to the vortex cores afresh, which are \emph{opposite} to initial ones, cf. Figs.~\ref{fig6} (a) and (c).
This phenomenon, the ``charge flipping'', was recently predicted to occur in a significantly different system, namely, vortices in nonlinear photonic lattices \cite{tristram}. The latter system does not conserve the angular momentum at all,
because the rotational symmetry is broken by the lattice. Nevertheless, the similarity with that system, which seems to be important for the effect to occur, is the presence of two interacting sub-systems which are given the freedom to exchange the angular momentum: the vortex and the lattice in Ref.~\cite{tristram}, and the two components of the vortex in the present case. It is commonly known from the studies of vortices in linear optics \cite{Soskin} that vortices in the phase fronts can annihilate or be born in pairs. The charge-flipping phenomenon introduces a new mechanism of such transformation in the nonlinear setting, through the exchange of the angular momentum between two nonlinearly (XPM) coupled subsystems. 

After the first charge flip, the components do not fully restore their annular shape (Fig.~\ref{fig6} (c)), and the maximum value of the partial spins which have the opposite signs, $S_2=-S_1\approx 0.83$ with $\psi \approx -0.462$, is attained at $z\approx 117300$  (Fig.~\ref{fig7}(a)). At the same time, the dipoles in both components start to rotate slowly (the angle $\varphi_0$ increases, see Fig.~\ref{fig7} (b)). We can introduce, therefore, the \textit{angular velocity} of the rotation, $d\varphi_0/dz$, as shown in Fig.~\ref{fig7}(c). However small, it demonstrates an important feature of the
correlation between the internal ``degrees of freedom" $\psi$ and $\varphi_0$. Indeed, for the exact stationary solutions, these two parameters are independent, while for the perturbed solutions in Figs.~\ref{fig6} and \ref{fig7}, they become coupled through
the growing instability modes. In particular, points where the angular velocity vanishes correspond to dipole-dipole soliton, with zero vorticity in both components, while fastest rotation is achieved when the partial spins in the components attain maximum absolute values.

\begin{figure}
\includegraphics[width=80mm]{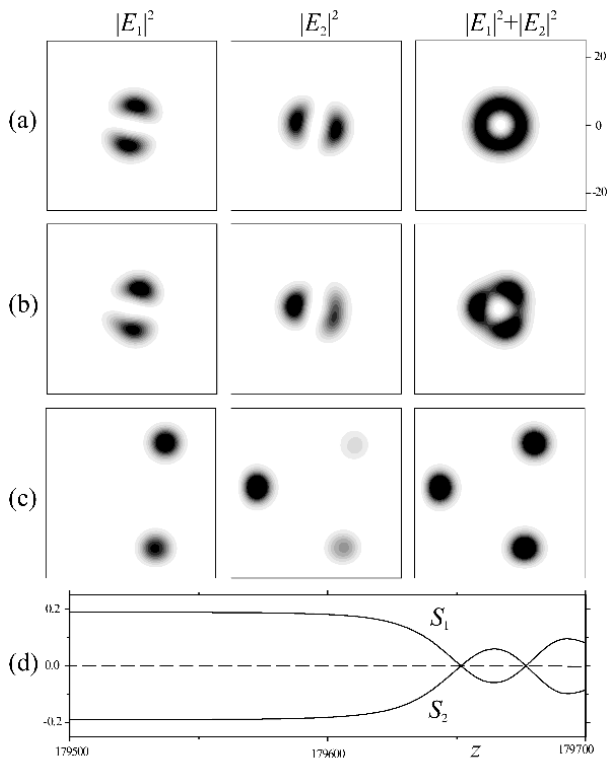}
\caption{Final stage with soliton ``explosion'' after quasi-stable evolution in Fig.~\ref{fig6} for the propagation distances $z=179500$ in (a), $179600$ in (b), and $179700$ in (c).}
\label{fig8}
\end{figure}

We continued the simulations, and eventually observed a break-up of the vortex soliton, as shown in Fig.~\ref{fig8}. It occurs within the distance of several hundreds of the propagation units, i.e. three orders of magnitude smaller than the previous stable
propagation, thus it may be considered as an ``explosion''. We have checked the evolution of the integral of motions at this stage, and, in particular, observed perfect conservation of the total spin, Fig.~\ref{fig8} (d), which rules out a numerical error as a probable cause of the ``explosion''. 

\begin{figure}
\includegraphics[width=70mm]{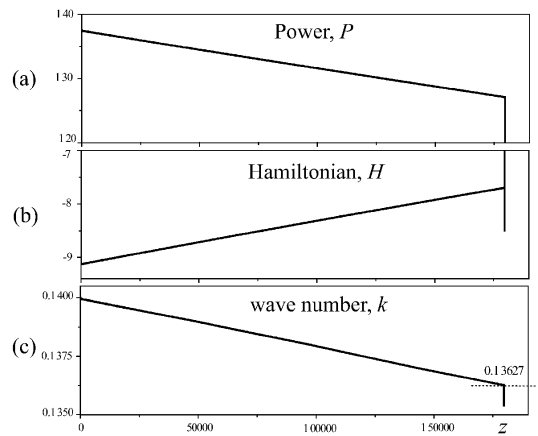}
\caption{Evolution of the total integral power (a), Hamiltonian (b), and accordingly defined propagation constant (c) in the course of the propagation shown in Figs.~\ref{fig6} and \ref{fig8}.} 
\label{fig9}
\end{figure}

In Fig.~\ref{fig9} we plot the total integral power and Hamiltonian, calculated in the course of the propagation, and notice small changes which naturally occurs due to the radiation emission from perturbed soliton. The breakup is accompanied by strong radiation which leads to sharp changes in the final segment of the diagrams. Using the relations displayed in Fig.~\ref{fig1}, we restore the corresponding value of the propagation constant $k$ and plot it in the Fig.~\ref{fig9} (c). As one can see, the propagation constant decreases, parallel to the power loss due to the radiation. The explosion occurs when the propagation constant reaches the value $k\approx 0.13627$, and the splitting follows the scenario already observed in Fig.~\ref{fig3} -- three splinters fly away along radial directions. We conclude that the sudden splitting of the otherwise ``intrinsically" unstable HV soliton happens because, as a result of the slow evolution of its integral characteristics due to the continuing radiation loss, it hits a boundary of the ``external" instability domain, after which it breaks apart. The latter boundary is found from the linear-stability analysis to be at $k=0.13582$ (see Table I), and we stress remarkable agreement and accuracy of the numerical procedure: both values coincide up to $10^{-3}$, $k\approx 0.136$. 

Deeper in the linear-stability domain, for instance for $k=0.16$, the HV soliton of the type $(1,-1)$ demonstrates no sign of instability for any reasonable propagation distance despite addition of initial noise, because the corresponding instability
mode with $s=2$ in Fig.~\ref{fig3} has a vanishingly small growth rate. We reflect the fact of small ``internal" instability by excluding corresponding domain from the final summary in Table I. 

We believe that the internal dynamics of the vectorial soliton reported above (Figs.~\ref{fig6} and~\ref{fig7}) for the particular case of $m=1$ and the CQ nonlinearity can manifest itself as a generic effect, for higher topological charges and in other systems, for example in the mixture of Bose-Einstein condensates \cite{BEC}.

\section{Instability of the vortex core close to cut-off}

\begin{figure}
\includegraphics[width=65mm]{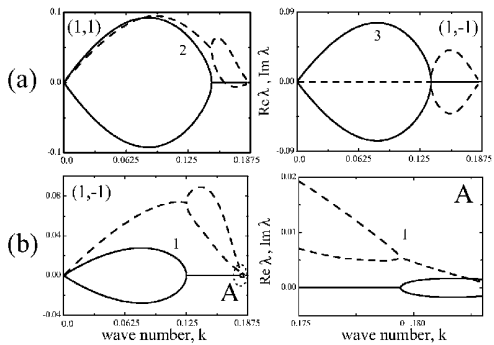}
\caption{(a) Real (solid) and imaginary (dashed) parts of the instability eigenvalues with the maximum growth rate for the vectorial solitons of the $(1,\pm 1)$ types. (b) An additional bifurcation (magnified in the inset A) of the perturbation mode with $s=1$, which occurs for the hidden-vorticity soliton of the $(+1,-1)$ type close to the cutoff. This instability gives rise to a shift of the soliton's core off the center.}
\label{fig10}
\end{figure}

As it was already stressed, Figs.~\ref{fig2} and \ref{fig4} demonstrate that, in contrast to the stability domain for the solitons of the $(m,m)$ type, the stability region for the HV solitons of the $(m,-m)$ type does not extend to the cutoff point. In this section, we aim to study the instability of the HV solitons close to this point. The simulations demonstrate that the corresponding instability mode with the azimuthal index $s=1$ leads to a shift of the vortex core. A possibility of this specific instability was earlier studied in Ref.~\cite{s1} for vortices in scalar models, using an analytical approximation for very broad annular solitons. It was concluded that such an instability may occur, but no linear unstable mode corresponding to the core shift was found [as before, the scalar case is exactly tantamount to the symmetric vectorial vortices of $(m,m)$ type considered here]. In contrast to that, Fig.~\ref{fig10} shows an explicit
example of such an instability mode for the HV solitons of the $(1,-1)$ type. 

\begin{figure}
\includegraphics[width=80mm]{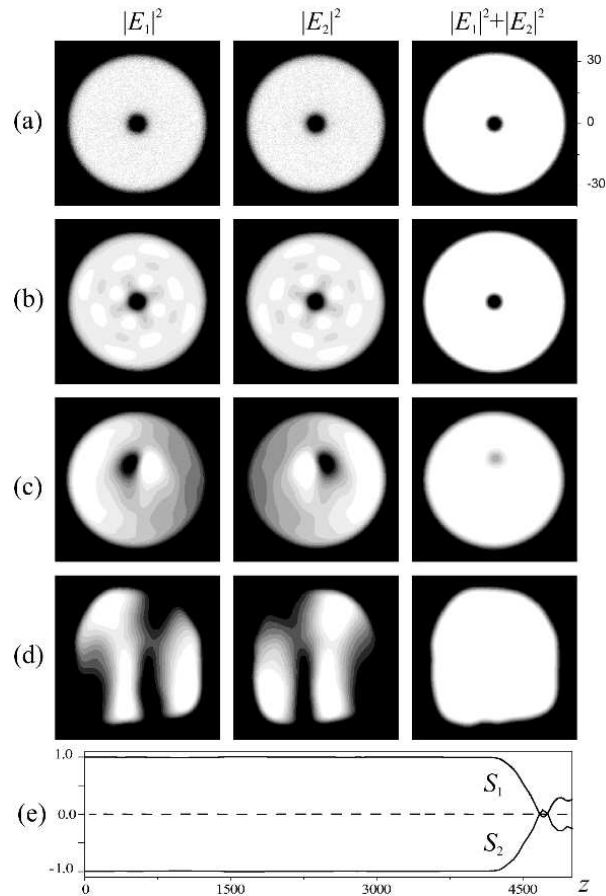}
\caption{Dynamics of the hidden-vorticity soliton of the $(1,-1)$ type with $k=0.18$ and initial $20$\% random-noise perturbation. The propagation distance is $z=0$ in (a), $500$ in (b), $4300$ in (c), and $5000$ in (d). As before, the frame (e) demonstrates the evolution of the partial spins (solid lines) in the two components and conservation of the total angular momentum (dashed line).}
\label{fig11}
\end{figure}

Numerical development of this instability is displayed for $k=0.18$ in Fig.~\ref{fig11}. Adding $20$\% of an initial random noise (Fig.~\ref{fig11} (a)) does not strongly affect the dynamics --- for up to $4000$ propagation units, it shows no sign of instability. The only visible action of the perturbation is excitation of the internal modes of the vectorial soliton, corresponding to the purely imaginary eigenvalues in the linear-perturbation spectrum (such eigenvalues are shown in Fig.~\ref{fig10} (a)). The pattern displayed in Fig.~\ref{fig11} (b) periodically repeats itself during the propagation. Similar long-lived internal modes have been recently observed in perturbed
evolution of scalar CQ vortex solitons \cite{internal}. It is noteworthy too that, for the symmetric vectorial vortex shown in Fig.~\ref{fig11}, the total intensity shows no sign of azimuthal modulation, as the intensity of the components sum up to the axially uniform distribution, similar to the case of ``intrinsic" instability shown in Fig.~\ref{fig6}. It is expected that the full set of internal modes of the vectorial vortex solitons should inherit all symmetry properties of the underlying stationary solutions; however, no verification of this assumption has been reported thus far.

At the distance $z=4300$, the growth of the unstable mode results in the separation of the dislocations in the two components, see Fig.~\ref{fig11} (c). The total-intensity distribution remains unmodulated, and the vortex core is invisible, because it is covered, in the total-intensity distribution, by the mutual displacement of the components: actually, an intensity maximum in one component lies on top of a minimum in the other. The shift of the vortex core quickly increases, and it moves to the outer
edge of the soliton. As a whole, the vectorial soliton remains localized, as is seen in Fig.~\ref{fig11} (d), and it possesses no vorticity, as the spin diagrams in Fig.~\ref{fig11} (e) demonstrate. Strong modulation of the components inside the soliton persists for the a long propagation distance after the vortex annihilation. We conclude that, in this case, the result of the development of the instability mode is establishment of a breather, i.e., a strongly perturbed and oscillating zero-spin soliton.

\section{Conclusion}

\begin{table}
\caption{Stability intervals and their size relative to the overall existence interval, $0<k<3/16\equiv 0.1875$, for the vectorial vortex solitons of sundry types.}
\begin{ruledtabular}
\begin{tabular}{lccr}
$(m,n)$& unstable      & stable            & \%\% \\
\hline\\
(1,1)  & $0<k<0.14855$ & $0.14855<k<0.18750$ &20.8\% \\
(2,2)  & $0<k<0.16190$ & $0.16190<k<0.18750$ &13.7\% \\
(3,3)  & $0<k<0.17005$ & $0.17005<k<0.18750$ &9.3\% \\
\hline\\
(1,-1) & $0<k<0.13582$ & $0.16163<k<0.17945$ &9.5\% \\
(2,-2) & $0<k<0.14884$ & $0.15620<k<0.15940$ &1.7\% \\
(3,-3) & $0<k<0.15866$ & $0.15866<k<0.15973$ &0.57\% \\
\end{tabular}
\end{ruledtabular}
\end{table}

We have demonstrated, for the first time, that 2D spatial solitons of the annular shape, carrying zero total vorticity, may be stable in the CQ medium, being supported by the hidden (implicit) vorticity, in quite a broad region. The output of the linear-stability analysis is summarized in Table I, where the stability domain and its size relative to the existence domain are shown. There is a single border between instability and stability regions for the explicit-vorticity solitons, of the $(m,m)$ type.
The stability domain for this type of the vectorial solitons extends up to the point of the transition to dark-soliton vortices (cutoff). For the implicit-vorticity solutions of $(m,-m)$ type, the situation is more complex. We identify the region of the relatively strong ``external" instability (shown in Table I), where unstable vectorial solitons with both implicit and explicit vorticity split into a set of fragments, the number of which is equal to the azimuthal index of the fastest growing mode of small perturbations. For larger $k$, and only for the hidden-vorticity solitons of the $(m,-m)$ type, there exist a domain of the very weak ``intrinsic" instability, where the vectorial soliton as a whole remains robust, while its components undergo a very slow internal evolution, ``sliding" through the solution family and exhibiting exchange of the angular momentum and charge flipping. In a vicinity of the cutoff, where the explicit vortices are stable, the hidden-vorticity solitons reveal a weak-instability mode that results in splitting of the phase dislocations in the two components, somewhat similar to the splitting of multiple-charged dark vortex solitons in the scalar model. Therefore, the stable solitons with hidden vorticity may be regarded as ``exceptionally bright" objects, unlike the familiar solitons carrying explicit vorticity, which always have stable dark-vortex counterparts. With the increase of the integer vorticity $m$, the stability regions of the vectorial solitons with both explicit and implicit vorticity quickly shrink.

\begin{acknowledgments}
A.S.D. gratefully acknowledges support from the Alexander von Humboldt Foundation. The work of B.A.M. was supported in a part by the Israel Science Foundation through the grant No. 8006/03. This author appreciate hospitality of the Nonlinear Physics Centre at the Research School of Physical Sciences and Engineering, Australian National University. \end{acknowledgments}

\end{document}